# Classification of IDS Alerts with Data Mining Techniques

Hany Nashat Gabra
Computer and Systems Engineering Department, Ain Shams University, Cairo, Egypt.
hanynashat@hotmail.com

Dr. Ayman M. Bahaa-Eldin
Computer and Systems Engineering Department, Ain Shams University, Cairo, Egypt.
ayman.bahaa@eng.asu.edu.eg

Prof. Huda Korashy
Computer and Systems Engineering Department, Ain Shams University, Cairo, Egypt.
hoda.korashy@eng.asu.edu.eg

## ABSTRACT

Intrusion detection systems (IDSs) have become a widely used measure for security systems. The main problem for those systems results is the irrelevant alerts on those results.
We will propose a data mining based method for classification to distinguish serious alerts and irrelevant one with a performance of 99.9 % which is better in comparison with the other recent data mining methods that have reached the performance of 97%. A ranked alerts list also created according to alert's importance to minimize human interventions.

**Keyword:** Intrusion Detection, Data Mining, Frequent Pattern, Frequent Itemset

## 1. INTRODUCTION

An IDS sensor can generate thousands of alerts in a day [1, 2]. Often vast majority of the alerts are false positives or of low importance [3,1]. More than 90% of those alerts being irrelevant [4, 5, 6] so an IDS alert log's analysis techniques are often used to distinguish the important IDS alerts from irrelevant events. Our result showed that the performance has been enhanced as we reduced the number of irrelevant alerts to 99.9 % in comparable with the performance of other recent techniques that reduced the number of irrelevant alerts by 74-97% only [4, 2, 7, 8, 1].

## 2. RELATED WORK

Data mining techniques first used for knowledge discovery from telecommunication event logs more than a decade ago [9]. Clifton and Gengo [10] have investigated the detection of frequent alert sequences and enhanced by Ferenc [11], Walter A. Kosters and Wim Pijls [12] this knowledge for creating IDS alert filters. Long et al [3] suggested a snort clustering algorithm. During the last 10 years, data mining based methods have also been proposed in many research papers [4, 5, 10, 3, 7, 8].

## 3. Mining Frequent Patterns

Mining frequent itemsets from a database has been solved largely by algorithms that are Apriori based and those that are pattern-tree growth techniques. Algorithms for mining of all existing techniques do not include generating frequent patterns for each transaction as needed for many applications.

**Table 1:** Example Alerts/items Data set Records

| Alerts | Item |
|---|---|
| Alert1 | 1 3 4 |
| Alert2 | 2 3 5 |
| Alert3 | 1 2 3 5 |

Assume a data set which contains alert records generated by an IDS system in Table 1 where the set of items I = {1, 2, 3, 16, 20} and the set of Alerts = {Alert1, Alert2, Alert3}.Mining all alerts that have similar frequent itemset at minimum support of 50% would require generating frequent itemsets with the alerts in the format [< itemset > Alert-list].

We proposed the AlertFp algorithm for mining frequent patterns with the Alerts where they occurred. Mining Fps with Alerts on an IDS log is an important goal of this algorithm where we are linking all frequent patterns to the alert transactions where they came from. Then count the number of frequent patter founded on each transaction. Finally all transactions in the dataset re-sorted according to the number of the related frequent patters. AlertFp algorithm represents each frequent k-pattern as form < $F_{k1}$, $Alert1_{k1}$, $Alert2_{k1}$, ..., $Alertm_{k1}$>, where $F_{k1}$ is the first frequent k-pattern, and $Alertm_{k1}$ is the mth Alert of the first frequent k-pattern. Thus, with this AlertFp technique the data set is scanned to obtain the candidate 1-itemsets with a list of their Alerts. The Alerts of each candidate pattern is implemented. Then, the count of each candidate pattern's Alerts is equivalent to the support of the pattern. After applying the frequent pattern mining algorithm to past IDS alert logs (AlertFp) to discover patterns that describe redundant alerts. Alert weight is measured by calculating Frequent Pattern Outlier Factor (FPOF) for each alert's transaction.

$$\text{FPOP}(t) = \frac{\sum_{X \subseteq t,\ X \in FPS(D, minisupport)} support(X)}{\|FPS(D, minisuport\ t)\|}$$

The interpretation of the above formula follows [19]. If a transaction t contains more frequent patterns, its FPOF value will be big, which indicates that it is unlikely to be an outlier.

In contrast, transactions with small FPOF values are likely to be outliers or to be considered as an interesting alert to be investigated by the security analyst.

By using $x \subseteq t, X \in FPS \sum_{(D, minisupport)} support(X)$s and re-order the IDS alerts by the simple FPOF for simplicity we will have the important alerts on the top of IDS log and irrelevant alerts will be pushed to the end of the log file.

**Algorithm 1**. (Alert:Computing Frequent Patterns with Alerts)
Algorithm AlertFp()
Input: A list of k-items, Alert Set of k-Alerts, mini-support s.
Output: A list of frequent patterns Fps and the relative Alert.
Begin
1. Scan the Data Set once to compute
2. Compute frequent pattern F1 from candidate k-itemsets
C1 as F1 = {list of k itemset with Alertslist count ≥ minsupport , Alert1counter}.
3. For Fi < k   i=1 m=0 Counter=0 do
Begin
3.1. If Fi € Alertmi then counter(m)++
3.2. i = i+1, m=m+1
3.3. Compute the next candidate set Ci+1 as F1
End

## 4. CASE STUDY

Snort [22] used in IDS sensor package that applies attack signatures for detecting suspicious network traffic and can emit alerts as syslog. Consider the below Snort sample (figure 1). This sample will be used to simply clarify the idea

```
7 1 508 WEB-MISC/doc/access  25 2 6/11/2010 8:57 AM 1136881320 2148203530 6 46,865 80
7 2 508 WEB-MISC/robots.txt/access 25 2 6/11/2010 8:57 AM 3632363311 2148203629 6 34,074 80
7 3 508 WEB-MISC/robots.txt/access 25 2 8/11/2010 8:59 AM 3632363313 2148203229 6 34,075 80
```

**Figure 1.** Snort alerts sample

The frequent patterns discovered from the sample IDS log as shown in figure 2.

```
* * * * *, t1, t2 ,t3                                                      Support: 3
 * * * * 25, t1, t2 ,t3                                                    Support: 3
7 * * * (25) , t1, t2 ,t3                                                  Support: 3
(7) * * * (25) 2, t1, t2 ,t3                                               Support: 3
(7) * 508 * (25) (2), t1, t2 ,t3                                           Support: 3
(7) * (508) * (25) (2) * * * * 6, t1, t2 ,t3                               Support: 3
(7) * (508) * (25) (2) * * * * (6) * 80, t1, t2 ,t3                        Support: 3
(7) * (508) WEB-MISC/robots.txt/access (25) (2) * * * * (6) * (80), t1, t2  Support: 2
(7) * (508) * (25) (2) * 8:57AM * * (6) * (80), t1, t2                     Support: 2
(7) * (508) * (25) (2) 6/11/2010 (8:57AM) * * (6) * (80) , t1, t2          Support: 2
```

**Figure 2.** Sample alert patterns

Finally the alerts are sorted in ascending order according to their simple FPOF and top p% of them is put into the set of candidate true alerts. The alerts are sorted in ascending order according to their weight (FPOF) as shown in figure 3.

```
simple FPOF t(3) = 7 ------------ 7 3 508 WEB-MISC/robots.txt/access 25 2
8/11/2010 8:59AM 3632363313 2148203229 6 34,075 80
simple FPOF t(1) = 8 ------------ 7 1 508 WEB-MISC/doc/access 25 2 6/11/2010
8:57AM 1136881320 2148203530 6 46,865 80
simple FPOF t(2) = 9 ------------ 7 2 508 WEB-MISC/robots.txt/access 25 2
6/11/2010 8:57AM 3632363311 2148203629 6 34,074 80
```

**Figure 3.** Output sample

## 5. IMPLEMENTATION AND PERFORMANCE

In this section, we describe our classifier implementation and experiments. In our setup, alerts sorted in a new separate log file for further review. Classifiers are rebuilt every midnight using the IDS sensor log data. Once the frequent pattern has been detected, it will be used for further alert classification. This allows for the classifier to adapt to new routine alert patterns with a reasonable learning time. The Outlier Factor will be calculated for each transaction, and then we will re-sort the transactions accordingly. In our experiments we have applied 5 artificial hacks from a specific source IP to be monitored on our result. Table 2 presents our experiment results on 22 June 2010 sample (with 28,670 records)

**Table 2:** Experimental results

| mini-support | frequent itemsets | attempted 5 attacks place | reduction |
|---|---|---|---|
| 2 | 101101522 | first 7 records | 99.975 % |
| 4 | 32589268 | first 24 records | 99.916 % |
| 6 | 23664252 | first 34 records | 99.882 % |

During the experiments, we measured the system reliability and accuracy (figure 4) for different support values comparable with the original attempted attacks and its place in the output file

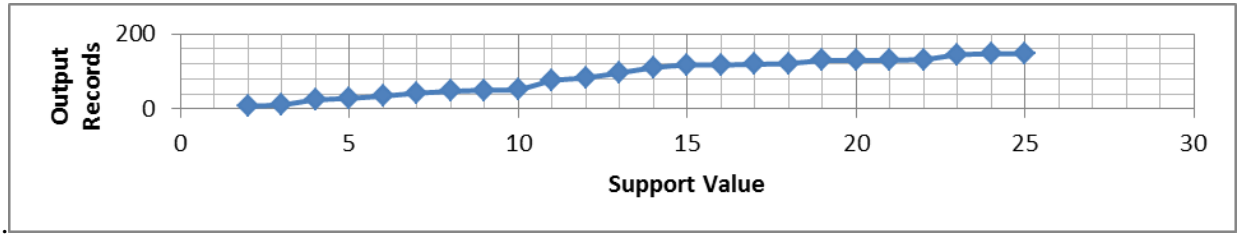

**Figure4.** mini support value vs. the 5 attacks in output

## 7. OPEN ISSUES AND FUTURE WORK

In this paper, we have presented a novel data mining based IDS alert classification method sorted for the security analysts according to the alert importance. Although our preliminary results are promising, one issue remains open – major changes in the arrival rate of routine alerts might be symptoms of large scale attacks, but are hard to detect. However, this is an inherent weakness of alert classification and sorting systems (e.g., see [6, 20, 13] for a related discussion). For the future work, we plan to research our classification method further, and study various statistical algorithms (e.g., time series analysis) for detecting unexpected fluctuations in the arrival rates of routine alerts.

## 8. REFERENCES

[1] Risto Vaarandi "Real-time classification of IDS alerts with data mining techniques," Proceedings of the 2009 IEEE MILCOM.
[2] J. Viinikka, H. Debar, L. Mé, A. Lehikoinen, and M. Tarvainen,"Processing intrusion detection alert aggregates with time series modeling", Information Fusion Journal 2009.
[3] J. Long, D. Schwartz, and S. Stoecklin."Distinguishing False from True Alerts in Snort by Data Mining Patterns of Alerts," in Proc. of 2006 SPIE Defense and Security Symposium, pp. 62410B-1--62410B-10.
[4] K. Julisch and M. Dacier. "Mining intrusion detection alarms for actionable knowledge" 2002 ACM SIGKDD Knowledge Discovery and Data Mining conference, pp. 366-375.
[5] K. Julisch. "Clustering Intrusion Detection Alarms to Support Root Cause Analysis," in ACM Transactions on Information and System Security, vol. 6(4), 2003, pp. 443-471.
[6] J. Viinikka, H. Debar, L. Mé, and R. Séguier. "Time Series Modeling for IDS Alert Management," in Proc. Of 2006 ACM Symposium on Information, Computer and Communications Security, pp. 102-113.
[7] S. O. Al-Mamory, H. Zhang, and A. R. Abbas. "IDS Alarms Reduction by Data

Mining," in Proc. of 2008 IEEE World Congress on Computational Intelligence, pp. 3564-3570.
[8] S. O. Al-Mamory and H. Zhang. "Intrusion Detection Alarms Reduction by Root Cause Analysis and Clustering" in Computer Communications, vol.32(2), 2009, pp. 419-430.
[9] K. Hätönen, M. Klemettinen, H. Mannila, P.Ronkainen, and H. Toivonen. in Proc. of 1996 International Conference on Data Engineering, pp. 115-122.
[10] C. Clifton and G. Gengo. "Developing Custom Intrusion Detection Filters Using Data Mining," in Proc. of 2000 MILCOM Symposium, pp. 440-443.
[11] Ferenc Bodon "fast APRIORI implementation" Informatics Laboratory, Computer and Automation Research Institute, Hungarian Academy of ciences", IEEE ICDM Workshop on Frequent Itemset Mining Implementations.
[12] Walter A. Kosters and Wim Pijls "Apriori, a depth-first implementation", volume 90 of CEUR Workshop Proceedings, CEUR-WS.org.
[13] J. Viinikka and H. Debar. "Monitoring IDS Background Noise Using EWMA Control Charts and Alert Information," in Proc. of 2004 RAID Symposium, pp. 166-187.
[14] B. Goethals."Frequent Pattern Mining" Technical Report, University of Helsinki 2002.
[15] M. J. Zaki and C.-J. Hsiao. "CHARM: An Efficient Algorithm for Closed Itemset Mining," in Proc. of 2002 SIAM International Conference on Data Mining, pp. 457- 473.
[16] T. Pietraszek. "Using Adaptive Alert Classification to Reduce False Positives in Intrusion Detection," in Proc. of 2004 RAID Symposium, pp. 102-124.
[17] L. Ertoz, E. Eilertson, A. Lazarevic, et al, "Detection of Novel Network Attacks Using Data Mining", Proceedings of DMSEC 2003, IEEE Press, New York, pp. 1-10, 2003.
[18] P. Dokas, L. Ertoz, V. Kumar, et al, "Data Mining for Network Intrusion Detection", AAAI/MIT Press, Cambridge, pp. 21-30, 2002.
[19] Z. He, X. Xu, J.Z. Huang, et al, "FP-Outlier: Frequent Pattern Based Outlier Detection", Computer Science and Information System, 2(1), pp. 103-118, 2005.
[20] J. Viinikka, H. Debar, L. Mé, A. Lehikoinen, and M. Tarvainen. "Processing intrusion detection alert aggregates with time series modeling," in Information Fusion Journal, 2009.
[21] Risto Vaarandi "Mining Event Logs with SLCT and LogHound", Proceedings of INTELLCOMM 2004: 293-308.
[22] Snort, http://www.snort.org/.